\documentstyle[floats,prd,aps,epsfig]{revtex}
\newcommand{\bce}{\begin{center}} 
\newcommand{\ece}{\end{center}}
\newcommand{\beq}{\begin{equation}} 
\newcommand{\eeq}{\end{equation}}
\newcommand{\beqy}{\begin{eqnarray}}
\newcommand{\eeqy}{\end{eqnarray}} 

\input epsf
\advance\voffset by 0.5in

\advance\voffset by -0.5in
\begin{document}
\title{Bremsstrahlung neutrinos from electron-electron scattering \\
       in a relativistic degenerate electron plasma}
\author{Prashanth Jaikumar\footnote{jaikumar@phy.anl.gov}$^1$, Charles Gale$^2$, and Dany Page$^3$}

\address{$^1$Physics Division, Argonne National Laboratory, Argonne IL 60439, USA.\\
         $^2$Department of Physics, McGill University, Montr\'{e}al QC H3A 2T8, Canada.\\
 	 $^3$Instituto de Astronom\'{\i}a, Universidad Nacional Aut\'onoma de M\'exico, 
             M\'exico D.F. 04510, Mexico.}

\maketitle
\begin{abstract}
We present a calculation of neutrino pair bremsstrahlung due to
electron-electron scattering in a relativistic degenerate plasma of
electrons. Proper treatment of the in-medium photon propagator, i.e.,
inclusion of Debye screening of the longitudinal part and Landau
damping of the transverse part, leads to a neutrino emissivity which
is several orders of magnitude larger than when Debye screening is
imposed for the tranverse part.  Our results show that this in-medium
process can compete with other sources of neutrino radiation and
can, in some cases, even be the dominant neutrino emission mechanism.
We also discuss the natural extension to quark-quark bremsstrahlung in
gapped and ungapped quark matter.  \\
\bigskip


\end{abstract}


\section{Introduction}
        \label{sec_intro}

For many astrophysical objects (e.g., supernovae or neutron stars) in
which high densities and/or temperatures are attained, energy losses
by neutrino emission are the driving force in their
evolution~\cite{RaffeltBook,YP04,PGW05}.  The simplest process of
$\beta$- and inverse $\beta$-decay of nucleons, the so-called direct
Urca process, leads to prodigious energy losses \cite{LPPH91} but is
possible only in cases where free nucleons are present and, even in
this case, can be kinematically forbidden.  Similar direct Urca
processes are also possible when hyperons~\cite{manju} or
quarks~\cite{IWA} are present and the associated emissivities dominate
the cooling, but the existence of these exotic phases is subject to
large uncertainties. In many cases, neutrinos can only be emitted
through higher order processes in which many-body effects in the
medium become important. A classic example is the plasmon decay
process of neutrino pair emission where the very existence of plasmons
is due to the medium's response to electromagnetic
perturbations. Neutrino radiation via bremsstrahlung comprises a large
subset of all neutrino emission processes where scattering of
particles, through electromagnetic or strong interactions, allows
transfer of energy-momentum accompanied by the emission of a Z$^0$
boson which immediately decays into a $\nu$-$\overline{\nu}$
pair. Typical and well studied examples are, among others, the case
of nucleon-nucleon, electron-ion and electron-phonon scattering in the neutron star's core and crust
(see, e.g., \cite{YKGH01} for a recent review).
However, neutrino pair bremsstrahlung emission in collisions of
relativistic particles has received little attention and
only a few attempts  exist in the literature, e.g., \cite{Haensel}. 
In this paper, we will be concerned with the importance of the medium 
in the modulation of the scattering interaction of relativistic electrons
through polarization effects, and its impact on the neutrino emissivity.

\vskip 0.2cm

In a QED plasma, static electric fields are screened while magnetic
fields suffer damping only at non-zero frequency.  For scattering in
the relativistic limit, when magnetic effects are important, soft
photon exchange can lead to a divergence from the unscreened magnetic
interaction.  To avoid this divergence, in early works on transport
coefficients in a QED plasma, Debye screening was imposed in both
polarization channels of the photon propagator in the covariant gauge,
even though it only applies to its longitudinal ``electric'' part. It
was later duly noted~\cite{BMPR90,HP93} that Landau damping of the
transverse ``magnetic'' part can generate a finite result in these
calculations. Landau damping instead of Debye screening of the
magnetic interaction implies higher scattering rates and hence smaller
thermal and electrical conductivities and shear viscosity \cite{HP93}.
Similarly, higher scattering rates should result in larger neutrino
emissivities for bremsstrahlung processes, and this is the issue we
address here by calculating the neutrino pair emission from
electron-electron bremsstrahlung in a degenerate relativistic electron
plasma.  Our results confirm these heuristic arguments and show that
erroneously assuming Debye screening of the magnetic interaction can
underestimate the neutrino emissivity by several orders of magnitude,
particularly at the low temperatures of relevance to a variety of
astrophysical environments.

\vskip 0.2cm

This paper is organized as follows: in section \ref{sec:many-body}, we
discuss the many-body effects that are relevant for the temperature
and density region under consideration and which must be taken into
account.  Section \ref{sec:brem} outlines the principal steps in the
calculation of the neutrino bremsstrahlung emissivity from
electron-electron collisions.  We present a comparison of the results
with other neutrino rates in three different contexts : from gapped
and ungapped quark matter in strange stars in section
\ref{sec:quarks}, from hot matter in a black-hole accretion torus in
section \ref{sec:BH}, and finally in the more conventional medium of a
neutron star, both in its crust and in its core, in section
\ref{sec:NS}. Our conclusions and outlook are presented in section
\ref{sec:Conclusion}. Some numerical details are explained in the
appendix.

\vskip 0.2cm
\section{Neutrino bremsstrahlung in electron-electron collisions}
         \label{sec:many-body}

We are interested primarily in the effects of Landau damping on the
bremsstrahlung rate since the screened case has already been
considered in~\cite{Haensel}.  A unified treatment, with consistent
inclusion of plasma effects, of neutrino bremsstrahlung in
electron-ion and electron-phonon collisions and plasmon decay in the
crust of neutron stars is described in~\cite{leinson}. Here, we derive
the emissivity for bremsstrahlung in scattering of relativistic
electrons, which is relevant to a host of astrophysical
conditions, and the results are also extended to quark-quark
bremsstrahlung, where the exchanged gluon is similarly dressed.
Neutrino bremsstrahlung is computed from the set of tree-level Feynman
graphs displayed in Fig. 1.  Before performing the explicit
computation, we discuss some many-body effects that can
potentially be important for our temperature and (electron or quark)
density regime.\\

\vskip 0.1cm

a) Landau damping: In the electron plasma, the exchanged photon is dressed by interaction with the electrons.  The longitudinal photon is Debye screened, while the transverse photon is Landau damped. At the one-loop level, an all-order resummation leads to a photon propagator that is modified from its ``bare'' expression as follows:
\beq
D^{\mu\nu}(\omega,k)=\frac{P_L^{\mu\nu}}{K^2-\Pi_L}+\frac{P_T^{\mu\nu}}{K^2-\Pi_T}\quad.   \label{prop1}
\eeq
where $K^2=\omega^2-k^2$ and $P_L^{\mu\nu}, P_T^{\mu\nu}$ are the longitudinal and transverse projection operators in~\cite{kap}. 
The longitudinal and transverse self-energies $\Pi_L$ and $\Pi_T$ are
in general complex functions of $\omega$ and $k$. These
self-energies can be obtained in the hard dense loop
approximation~\footnote{We are ignoring the smaller contribution from
finite temperature effects (hard thermal loops) since they are small for a degenerate QED plasma.}.  Since the
exchanged photon also has small energy, we will use approximate
forms~\cite{kap} in the nearly static limit $\omega/k\ll 1$:
\beq
\Pi_L=m_D^2 \quad {\rm and} \quad 
\Pi_T=m_M^2=i\left(\frac{\pi}{4}\right)\frac{\omega}{k}m_D^2\quad.
\eeq    
For degenerate electrons, characterized by a chemical potential $\mu_e$,
\beq 
m_D^2=e^2\mu_e^2/3 \quad,  
\eeq 
 but for the sake of generality, we use $m_D$ to denote the
 phase-dependent screening mass. Its explicit expression is to be
 understood by the particular case under consideration.

\vskip 0.2cm

b) Vertex renormalization: Neutrinos are emitted with average energies that are thermal, and in the degenerate regime $T/\mu_e\ll 1$. This implies that their wavenumber $k_{\nu}$ is such that $k_{\nu}^2/m_D^2\ll 1$. Physically, this means that at sufficiently low temperatures, and large electron densities, the neutrino's wavelength is comparable to or exceeds the Debye charge radius. In this case, as shown by Leinson~\cite{leinson}, the effective weak charge seen by the neutrino is renormalized. The correction factor due to this effect can be absorbed simply into ``effective'' electroweak couplings, and we take this into account in our calculations. 

\vskip 0.2cm

c) Non-Fermi liquid effects: In normal quark matter with $N_f$ quark flavors at a common chemical potential $\mu$, the absence of screening for the chromomagnetic gluon leads to a breakdown of Fermi liquid behaviour for quarks at a scale $T\sim m~{\rm exp}(-9\pi/4\alpha_c)$~\cite{SS1} where $m^2=N_f\alpha_c\mu^2/\pi$ is the electric screening mass, with $\alpha_c=g^2/4\pi$ where $g$ is the QCD coupling constant. The neutrino emission rate for the quark direct urca process is consequently enhanced~\cite{SS2}. This effect applies also to the electron plasma, but we ignore it here since the smallness of the QED coupling constant implies that this effect sets in only at exponentially small temperatures.

\vskip 0.2cm

d) LPM effect: Suppression of soft radiation due to multiple scatterings between electrons (the Landau-Pomeranchuk-Migdal or LPM effect) becomes important when the formation time $t_f$ of the neutrino wave-packet is comparable to the mean collision time $t_c$ between electrons. The ratio $t_f/t_c$, for a degenerate plasma ($T\ll\mu_e$) behaves approximately as $\sqrt{\alpha}T/\mu_e$ if only screening is considered, and as $\alpha$ when damping is taken into account ($\alpha$ is the fine structure constant). In either case, the ratio is much smaller than 1, so that an inclusion of the LPM effect would make only a 1\% correction to our results. The large degeneracy at the Fermi surface implies that the mean time between electron-electron collisions is much greater than the formation time of the neutrino wave-packet.

\begin{figure}[t]  
\bce
\epsfig{file=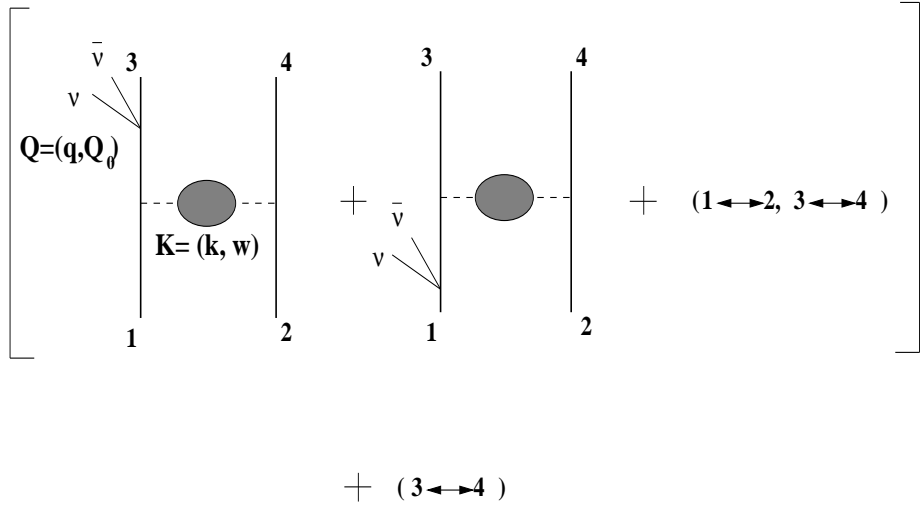,height=8.0cm,width=14.0cm}
\ece
\caption{The Feynman diagrams for neutrino bremsstrahlung in electron-electron scattering. The first 4 diagrams, enclosed within square brackets, are the ``direct'' contributions while the interchange ($3\leftrightarrow 4$) generates the ``exchange'' diagrams. The blob represents the one-loop resummed photon propagator (see text for details).}
\label{figure1}
\end{figure}

\section{Neutrino Emissivity from the bremsstrahlung process}
         \label{sec:brem}

The neutrino emissivity from the bremsstrahlung process is
\beqy
\label{emiss1} 
 Q_{\rm ee-Br} = \frac{2\pi}{\hbar}&&\biggl[\prod_{i=1}^{4}\int\frac{d^3p_i}{(2\pi)^32E_i}\biggr]\int\frac{d^3q_{1}}{(2\pi)^32\omega_{1}}\frac{d^3q_{2}}{(2\pi)^32\omega_{2}}Q_0\sum_{spin}\frac{|M|^2}{s}\\  \nonumber
&\times&n_F(E_1)n_F(E_2)\tilde{n}_F(E_3)\tilde{n}_F(E_4)(2\pi)^3\delta^3
({\bf P_f-P_i})\delta(E_{\bf f}-E_{\bf i})\quad. 
\eeqy
The subscripts $i=1$ to 4 on $p$ refer to electrons (1,2 label the
incoming   states and 3,4 the outgoing states), while {\bf i,f} refer
to total initial and final three-momentum/energy accordingly.
$Q_{1}=(\omega_{1},q_1),Q_{2}=(\omega_{2},q_{2})$ are respectively the
neutrino and anti-neutrino four-momenta with
$Q=(Q_0,q)=(\omega_{1}+\omega_{2},q_1+q_2)$. The symmetry factor
$s=2$. The phase space for electrons is convolved with the
appropriate Fermi distribution functions $n_F(E_i)=1/({\rm
e}^{(E_i-\mu_e)/T}+1)$ (in $k_B=1$ units) and ${\tilde n_F}=1-n_F$,
respectively. The matrix element $M$ is obtained by the application of
Feynman rules to the diagrams shown in Fig.~\ref{figure1}. Since
we work in the degenerate regime, the emissivity is dominated by
contributions from low energy neutrinos (typically of order the
temperature $T$, where $T\ll \mu_e$) and Low's theorem~\cite{Low58}
may be used to simplify the calculation of $Q_{\rm ee-Br}$. As a
consequence, the propagator of the internal fermion line can be
approximated as
\beq S_e(p\pm Q)=\pm\frac{1}{Q_0} \quad .
\eeq 
A further approximation can be made by noting that interference
between the direct and exchange diagrams is strongly suppressed
relative to their squares taken individually, and hence may be
neglected~\cite{Haensel}. Then, in the low-energy limit, the emission from the
exchange diagrams simply doubles the overall result for the emissivity
from the direct diagrams. Since there is a symmetry factor of $s=2$ in
the denominator of the emissivity expression eqn.~(\ref{emiss1}), we
may ignore the identity of the particles altogether and evaluate the
emissivity from the direct diagrams alone (see caption Fig.~\ref{figure1}).
\vskip 0.2cm

Aside from the equation for the dressed photon propagator (eqn.(\ref{prop1})),
we also require the weak interaction Hamiltonian (in the 4-Fermi form) given by
\beqy
H_{4f}&=&\frac{G_F}{2\sqrt{2}}\Gamma_{\lambda}\bar{\psi}_{\nu}\gamma^{\lambda}(1-\gamma_5)\psi_{\nu}\quad,\nonumber \\
\Gamma_{\lambda}&=&\bar{\psi}_e\gamma_{\lambda}(c_{V}-\gamma_5c_{A})\psi_e\quad .
\eeqy
The weak vector and axial-vector couplings are
\beqy
c_V^e&=&1+4{\rm sin}^2\theta_W, \quad c_A^e=1 \quad .\nonumber \\
c_V^{\mu,\tau}&=&-1+4{\rm sin}^2\theta_W, \quad c_A^{\mu,\tau}=-1\quad .
\eeqy
where the Weinberg angle ${\rm sin}^2\theta_W=0.231$. Neutrino-pairs of all three flavors ($e,\mu,\tau$) contribute incoherently to the emissivity, so that we may define a ``total'' squared coupling constant $c_+^2=\sum_{\nu}((c_V^{\nu})^2+(c_A^{\nu})^2)=6.696$. The vertex renormalization effect discussed in the previous section introduces a suppression factor $s=0.449$. This implies a total ``effective'' squared coupling constant $c_{\rm eff}^2=3.006$. We will use the symbols $c_V$ and $c_A$ for the following calculation, and substitute the effective coupling constant only at the end. Evaluating the spin-summed squared matrix element with the low-energy approximations for neutrinos, and utilizing symmetries between the electron ``labels'' (1,2,3,4), we obtain (neglecting terms of 
{${\cal O}(m_e^2/\mu_e^2)$)
\beqy
\frac{|M|^2}{s}&=&\frac{256e^4G_F^2}{Q_0^2}L(Q_{1},Q_{2})^{\rho\lambda}D^{\alpha\mu}D^{\beta\nu}p_{2\alpha}p_{2\beta}\times[ic_Vc_A\epsilon_{\nu\lambda}^{\gamma\delta}(T1)_{\gamma\delta\mu\rho}+(c_V^2+c_A^2)(T2)_{\nu\lambda\mu\rho}] \quad,\label{msquare}\\
(T1)_{\gamma\delta\mu\rho}&=&[\left\{g_{\delta\rho}(p_{1\mu}p_{3\gamma}+1\leftrightarrow 3)+g_{\gamma\mu}(p_{1\rho}p_{3\delta}+1\leftrightarrow 3)+p_1\cdot p_3(g_{\gamma\rho}g_{\mu\delta})\right\}-(\gamma\leftrightarrow\delta)] \quad , \\
L(Q_{1},Q_{2})^{\rho\lambda}&=&8(Q_1^{\lambda}Q_2^{\rho}+Q_1^{\rho}Q_2^{\lambda}-g^{\lambda\rho}Q_1\cdot Q_2-i\epsilon^{\mu^{\prime}\nu^{\prime}\lambda\rho}Q_{1\mu^{\prime}}Q_{2\nu^{\prime}}) \quad .
\eeqy
All vectors appearing in the above equation are 4-vectors, and $T2$ is obtained from $T1$ with the interchange of indices $(\delta\leftrightarrow\lambda, \gamma\leftrightarrow\nu)$. The neutrino phase space integrals are easily performed by introducing a delta function 
as 
\beq
1 = \int d^4Q~\delta^4(Q-Q_{1}-Q_{2})\,,
\eeq
and then using Lenard's identity
\beq
N_{\alpha\beta}=\int
\hspace{0.05in} \frac{d^3Q_{1}}{2\omega_{1}} \frac{d^3Q_{2}}{2\omega_{2}}~
Q_{1\alpha}Q_{2\beta}
\hspace{0.05in} \delta^{4}(Q-Q_{1} - Q_{2})=\frac{\pi
}{24}(Q^{2}g_{\alpha\beta} +
2Q_{\alpha}Q_{\beta})\Theta(Q_{0})\Theta(Q_{0}^{2} - {\bf q}^2) \quad.
\eeq 

Before performing further contractions, we choose to retain only the
transverse part of the photon propagator in eqn.(\ref{msquare}). We
will justify this {\it a posteriori} when the contribution from Landau
damping is found to be much larger than naive
screening~\cite{Haensel}. We focus on this effect, which is expected
to be important due to the exchange of soft photons. The remaining
tensor contractions in eqn.(\ref{msquare}) finally yield
\beqy Q_{\rm ee-Br}&=&A\int dQ_0Q_0^4P \,,\label{AP}\\
A&=&\left(\frac{16}{45}\right)\frac{1}{(2\pi)^{12}}\frac{e^4G_F^2(c_V^2+c_A^2)}{\hbar~p_{F_e}^4}\,,\\
P&=&\left[\int\prod_{i=1}^{4}d^3p_i\right]n_{F}(E_1)n_{F}(E_2){\tilde
n}_{F}(E_3){\tilde n}_{F}(E_4)\delta^3   ({\bf
P_f-P_i})\delta(Q_0+E_{\bf f}-E_{\bf i}) \nonumber \\
&&\times\frac{1}{|K^2-m_M^2|^2}\times\left\{|{\bf p}_2\times\hat{\bf
k}|^2~(13E_1E_3-{\bf p}_1\cdot{\bf p}_3)\right\}\quad. \label{AP2}
\eeqy 
where boldface quantities denote 3-vectors. The momentum
integrals over $d^3p_2,d^3p_3$ are performed first using the techniques
described in~\cite{HP93}. Since the electron mass $m_e\ll p_{F_e}$, the
electron Fermi momentum, the electron energies $E_i$ may be traded
for the momentum magnitudes $p_i$ at the Fermi surface. Then $P$ in eqn.(\ref{AP2}) becomes 
\beqy
P&=&24\pi^2\int_{-\infty}^{\infty}d\omega\int\frac{d^3k}{k^2|K^2-m_M^2|^2}\Theta(k-|\omega|)\int_{\frac{k+\omega}{2}}^{\infty}dp_4p_4(p_4-\omega)^3{\rm
sin}^2\theta_0\tilde{n}_F(p_4)n_F(p_4-\omega)\nonumber\\
&&\times\biggl\{\theta(\omega)\theta(k-(\omega+Q_0))+\theta(-\omega)\left[\theta(Q_0-|\omega|)\theta(k-(Q_0-|\omega|))+\theta(|\omega|-Q_0)\theta(k-(|\omega|-Q_0))\right]\biggr\}\nonumber\\
&&\times\int_{\frac{k+\omega+Q_0}{2}}^{\infty}dp_1p_1(p_1-\omega-Q_0)^3n_F(p_1)\tilde{n}_F(p_1-\omega-Q_0)
\label{Pexp} \,,\\ {\rm
cos}~{\theta_0}&=&\left(\frac{k^2-\omega^2+2p_4\omega}{2p_4k}\right)\quad .
\eeqy 
Next, the $dp_1,dp_4$ integrals are performed, which fixes the
limits on the $dk$ and $d\omega$ integrations. At this stage, it is
convenient to perform the $d^3k$ integral, noting that ${\rm
sin}^2\theta_0$ is a function of $\omega,k$. Explicitly, this integral
is of the form 
\beq I_k=\int_{l_1}^{2\mu_e}~dk\frac{{\rm
sin}^2\theta_0}{(\omega^2-k^2)^2+m_M^4}\quad,\label{kintegral} 
\eeq 
where
$l_1\sim T$ is the lower integration limit as determined by the
$\Theta$-functions involving $k$. The evaluation of this integral is
detailed in appendix A. The result is 
\beq 
I_k\approx \frac{4}{3\pi
m_D^2\omega}\left[\frac{\pi}{2}-{\rm
tan}^{-1}\left(\frac{l_1^3}{\frac{\pi}{4}m_D^2\omega}\right)\right] \quad .
\eeq
We restrict our temperature and density range such that this
approximation makes only a 10\% error in the final analytic expression
for the emissivity, which is acceptable. Using this approximation, $P$
reduces to 
\beqy 
\label{Pexp2}
P&=&\frac{128\pi^2}{m_D^2}\mu_e^8T^2N(\alpha),\\ 
N(\alpha)&=&\int_0^{\infty}d\bar{\Omega}\bar{\Omega}^4{\rm
e}^{-\bar{\Omega}}I(\bar{\Omega}) ;\quad \bar{\Omega}=\frac{Q_0}{2T}\,,\quad \alpha=\frac{m_D}{2T}\quad.\label{nalpha}
\eeqy 
where $I(\bar{\Omega})$ is a smooth function of
$\bar{\Omega}$ obtained upon integration over $\omega$ (see
Appendix B for details).  Substituting the resultant expression for
$P$ from eqn.(\ref{Pexp2}) in eqn.(\ref{AP}), we are left with an
overall integral over $Q_0$, which is numerically tractable. The final
result for the neutrino emissivity reads (we have chosen natural units $\hbar=c=1$; a factor of $\hbar^{10}c^9$ should be inserted in the denominator to recover ``standard'' units)
\beq 
Q_{\rm ee-Br}=\left(\frac{16}{45\pi^{10}}\right)\frac{e^4G_F^2c_{\rm eff}^2
T^7\mu_e^4}{m_D^2}N(\alpha)\quad. \label{bremss1}\\ 
\eeq
Substituting the values of the physical constants, and defining
$T_9$ as the temperature in units of $10^9$K, $\mu_{10}$ as the
electron chemical potential in units of 10 MeV, we obtain 
\beq 
Q_{\rm ee-Br}=4.6\times 10^{16} \; T_9^7 \; \mu_{10}^2 \; \tilde{N}(\alpha) ~{\rm
erg~cm}^{-3}{~\rm s}^{-1}\quad. \label{bremss2} 
\eeq 
where we also introduced a normalized $\tilde{N}(\alpha) = N(\alpha)/212$
such that $\tilde{N}(\alpha) \leq 1$ (See Figure~\ref{figure5}).
The emissivity is plotted as a function of temperature for an electron chemical potential $\mu_e=30$ MeV in Fig.\ref{figure2}. For
comparison, the emissivity as computed from eqn.(51) of
ref.~\cite{Haensel}, which assumed Debye screening of both the 
longitudinal and transverse parts of the photon propagator, is also
shown. The inclusion of Landau damping has modified the temperature
and density dependence of the emissivity from the case of pure
screening. In our case, the emissivity scales as $\mu_e^2~T^7$ while
in the screened case, it behaves as $\mu_e~T^8$. Note that in our
result, the quantity $N(\alpha)$ is also temperature dependent, so
that a slight curvature in the log-plot of emissivity versus
temperature is to be expected. In case of~\cite{Haensel}, a similar
numerical integral appears but is temperature-independent and their
result yields a straight line. The $T_9^7$ dependence implies a
slower fall-off of the emissivity with decreasing temperature,
reflecting the increased importance of soft photon exchange at lower
temperatures. This is due to the fact that electrons have smaller
thermal energies, and cannot scatter far from the Fermi surface. The
soft exchanges are not screened, and this leads to the enhancement of
the emissivity, when compared to previous results which did not
incorporate the Landau damping. At high temperatures ($T_9\gg 10$ and beyond), 
the electrons start to become non-degenerate, and our approximations would break down. Nevertheless, the convergence of our result to that of ref.~\cite{Haensel} with increasing temperature is expected since both must approach the emissivity computed in the non-degenerate case~\cite{CAZ}. 
 
\begin{figure}[!ht]
\bce
\epsfig{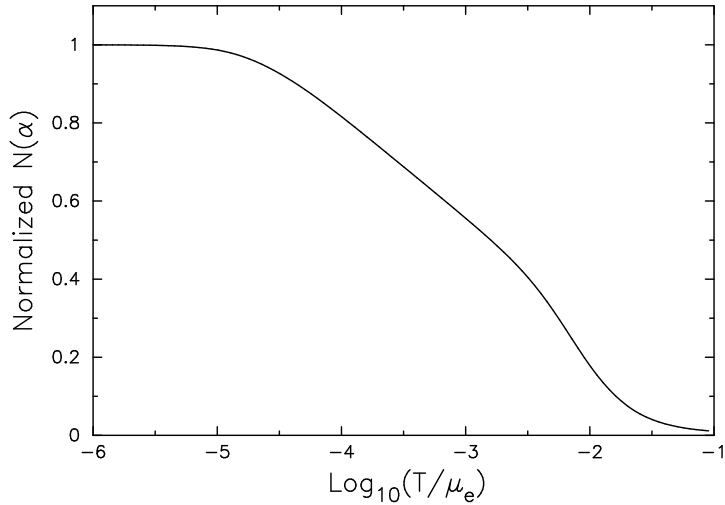}
\ece
\caption{The normalized $\tilde{N}(\alpha) \equiv N(\alpha)/212$ from eqn.(\protect\ref{nalpha}) }
\label{figure5}
\end{figure}

\vskip 0.2cm 

\begin{figure}[!ht]
\bce
\epsfig{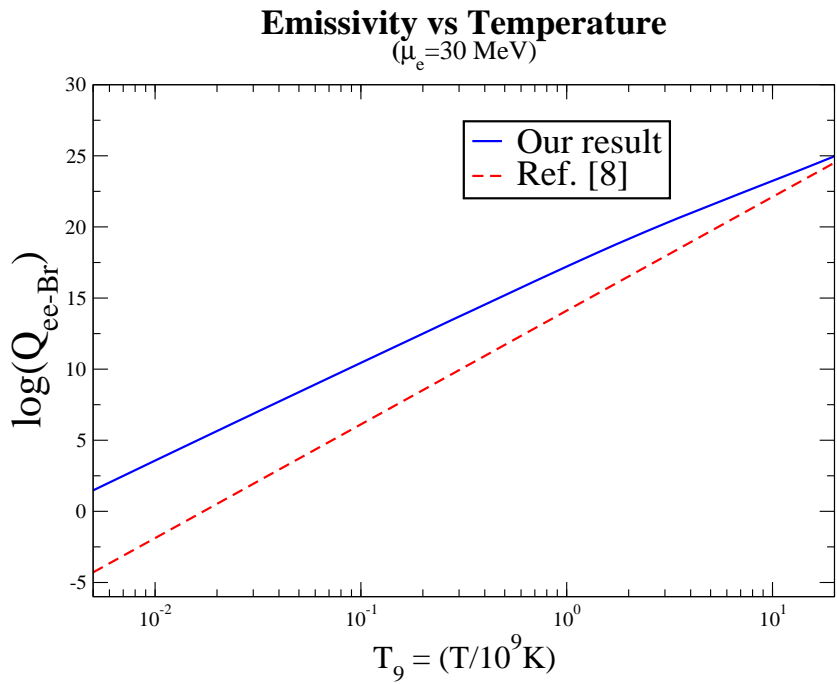}
\ece
\caption []{The neutrino emissivity (on a log scale) from eqn.(\ref{bremss2}) of this paper and eqn.(51) of~\cite{Haensel}. The chemical potential is fixed at $\mu_e=30$ MeV.}
\label{figure2}
\end{figure}
\vskip 0.2cm

There are other sources of neutrino emission from a QED plasma, such as the plasmon process~\cite{Bra,rat}, where an in-medium photon dressed by the electron plasma can decay into $\nu\bar{\nu}$ pairs, and the plasma photo-neutrino process~\cite{rat2} $e^{\pm}+\gamma\rightarrow e^{\pm}+\nu\bar{\nu}$. Formally, the plasmon process can interfere with $\nu\bar{\nu}$ bremsstrahlung when the exchanged photon in electron-electron collisions is almost on-shell. This is because the lifetime of such a photon is very long and it can be regarded as a valid in-medium excitation just like a plasmon. However, these two contributions can be disentangled when the temperature and density are such that only one of them dominates~\cite{Perez}. Studies of the plasmon process have shown that for a degenerate electron plasma at low temperatures such that $T\ll\omega_P$, where $\omega_P$ is the plasma frequency, the total emissivity acquires an exponential factor ${\rm e}^{-\omega_P/T}$. For example, the emissivity from the decay of the transverse mode (photon) is~\footnote{The emissivity can be decomposed into longitudinal and transverse parts for the vector channel, and an axial channel. They are all exponentially suppressed as ${\rm exp}(-\omega_P/T)$ though the pre-factors are different~\cite{rat}.}~\cite{rat}  
\beq
Q_T\simeq\frac{c_V^2G_F^2}{48\pi^4\alpha}\sqrt{\frac{\pi}{2}}\omega_P^{15/2}T^{3/2}{\rm e}^{-\omega_P/T}\quad .
\eeq
The plasma frequency $\omega_P$ for a relativistic degenerate plasma  is a function of the electron chemical potential
\beq
\label{omegap}
\omega_P^2\cong\frac{4\alpha}{3\pi}
\mu_e^2\biggl(1+\frac{\pi^2T^2}{3\mu_e^2}\biggr) \,.
\eeq
For $\omega_P\sim{\cal O}({\rm MeV})$ and for our temperature regime ($T\leq 1$ MeV), the plasmon decay rate and the plasma photo-neutrino rate are
strongly suppressed, while the bremsstrahlung rate is not.
In the following sections, we apply our 
main result, eqn.(\ref{bremss2}), to explore and compare neutrino rates in
a variety of astrophysical environments.

\section{Comparison to neutrino rates in strange stars}
         \label{sec:quarks}

Self-bound quark matter made of up, down and strange quarks can exist
at high density, as evidenced by model calculations~\cite{Farhi}. 
An entire star made of this phase is known as a strange star. Having obtained a modified in-medium neutrino radiation rate, we would
like to identify the physical conditions under which neutrino emission
from relativistic electron-electron scattering, and quark-quark scattering,
can affect strange star cooling.

\subsection{The strange star surface}

In case the star's surface is not covered by baryonic matter, it has a
bare quark surface covered by a thin layer of degenerate electrons
\cite{AFO86}, termed the electrosphere. It has recently been suggested
that the bare quark surface is not as sharp as previously thought but
may rather consist of a crust of quark nuggets embedded in a
degenerate electron gas~\cite{JRS05}. These strange nuggets have
larger electric charge than nuclei in a nuclear crust, but are very
sparsely distributed, so that neutrino luminosity from electron-nugget
bremsstrahlung will still be less than that from electron-electron
bremsstrahlung, as calculated in this paper. Another intriguing
scenario develops if the underlying quark matter is in a gapped phase,
since the neutrino emissivity from quarks is severely attenuated at
temperatures $T\ll T_c$ where $T_c$ is the critical temperature for
pairing. The specific heat is similarly affected if electrons are not
admixed with quark matter (as is the case with the color-flavor-locked
or CFL phase~\cite{Krish}). Despite the enforced charge neutrality of
the CFL phase in bulk, surface effects still lead to the formation of
a degenerate electron plasma at the star's surface~\cite{M03-U04} and
a possible nugget envelope along the lines of~\cite{JRS05} may be
present. Neutrinos from electron-electron bremsstrahlung in the
envelope will then be the dominant energy sink for a strange star composed
of stable CFL matter all the way up to its surface. In addition, the
superfluid nature of the CFL phase implies a large thermal
conductivity~\cite{Shovy}, so that neutrino cooling is determined by
the neutrino emission properties of the surface~\cite{jps}. Below,
we examine the dominant sources for neutrino emission in the bulk of the strange star, since the true phase of quark matter at supra-nuclear density remains as yet unknown, and make a comparison to the surface or envelope emission rates.

\subsection{Ungapped quark matter}

If quarks of all flavors are ungapped, the direct urca process (weak equilibrium between $d$ and $u$ quarks) is dominant, with~\cite{Iwamoto}
\beq
Q_{\rm durca}=8.8\times 10^{26}~\alpha_c\left(\frac{n_B}{n_0}\right) Y_e^{1/3}T_9^6~{\rm erg~cm}^{-3}{\rm s}^{-1}\quad . \label{durca}
\eeq 
With typical values $\alpha_c=0.1$, baryon density $n_B=2n_0$ ($n_0=0.16$~fm$^{-3}$ is the saturation density of nuclear matter) and electron fraction $Y_e=10^{-5}$, a comparison of eqn.(\ref{durca}) with eqn.(\ref{bremss2}) shows that the former dominates by more than 8 orders of magnitude for the entire temperature range under consideration. With the inclusion of the non-Fermi liquid effects in ungapped quark matter, the emissivity is~\cite{SS2}
\beq
Q=Q_{\rm durca}\left[1+0.014~{\rm ln}\left(\frac{235.85}{T_9}\left(\frac{n_B}{n_0}\right)^{1/3}\right)\right]^2~{\rm erg~cm}^{-3}{\rm s}^{-1}\quad . \label{nfdurca}
\eeq
Other neutrino emission processes from ungapped quark matter, such as from quark-quark scattering, also exceed the rate computed here. Therefore, the presence of ungapped quark matter below the surface obviates a consideration of neutrino emission from the surface, at least as regards the cooling behaviour of the star.

\subsection{Gapped quark matter}

In the CFL phase, all 9 quarks are gapped, and the expectation is that neutrino emission processes involving free quarks are suppressed by the energy cost to be paid in exciting them above the gap $\Delta\gg T$. This is not the complete story, however, since the CFL phase supports Goldstone modes (8 massive and 1 massless) that require much less energy to excite. Importantly, they can all couple to neutrinos, and contribute to the emissivity. Among them, the most important processes are the electroweak decay of the kaon, and of the exactly massless $H$ boson through a momentum dependent coupling to neutrinos~\cite{RST}. The latter has a very weak temperature dependence~\cite{jps} and can be effectively ignored until very late times in the star's history. A comparison of the neutrino rate from kaon decay (eqn.(30) in ref.~\cite{jps}) with eqn(\ref{bremss2}) is shown in Fig.\ref{cflcomp}, revealing that surface neutrino emission from the electrosphere starts to dominate for temperatures less than $8\times 10^8$K or so. The reason for the sharp fall-off of the neutrino emissivity in the CFL phase is the Boltzmann factor of ${\rm exp}(-m_{GB}/T)$ where $m_{GB}$, the mass of the Goldstone boson is typically 10 MeV.

\vskip 0.2cm 

\begin{figure}[!ht]
\bce
\epsfig{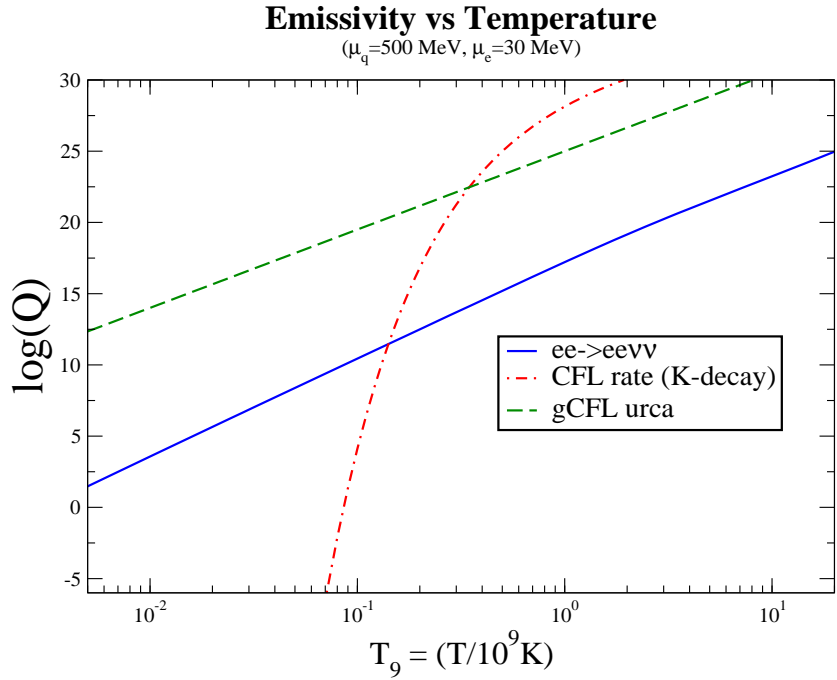}
\ece
\caption[]{Comparison of neutrino emissivity (on a log scale) from eqn.(\ref{bremss2})(solid) with eqn.(30) of ref.~\cite{jps}(dashed), and eqn.(31) of ref.~\cite{KR04}(dash-dotted). The quark chemical potential is fixed at $\mu_q=500$ MeV and the electron chemical potential at $\mu_e=30$ MeV.}
\label{cflcomp}
\end{figure}

Stresses placed on the CFL phase by virtue of a finite strange quark
mass can trigger the formation of new ground states. For example, the
lightest massive boson can Bose-condense, leading to neutral kaon
condensation (the ${\rm CFLK}^0$ phase)~\cite{TS1,TS2,PB}. While the
decreasing mass can initially increase the emissivity from the decay
$K^0\rightarrow \nu\bar{\nu}$~\footnote{At first sight, such a process
violates helicity conservation, but the presence of a preferred frame implies
that the decay is indeed allowed in a frame moving with respect to the
medium~\cite{jps}.}, phase space considerations eventually shut off
neutrino emission. The dominant rates come from processes such as
$K^+\rightarrow e^+{\nu_e}$ (and the analog for the $K^-$) and
$K^+e^-\rightarrow \nu_e$, which can occur due to the dispersion
relation turning space-like beyond a certain momentum. The associated
neutrino rates are enhanced over the CFL rates, but still suffer from
exponential suppression due to Boltzmann factors, and cannot compete
with the surface emission when the temperature drops below 0.1 MeV or
so.

\vskip 0.2cm

The CFL and CFL${\rm K}^{0}$ phases are the likely ground state of
dense quark matter only at extremely high quark chemical potential. At
densities relevant to neutron stars, with $\mu_q\sim 500$ MeV or less,
and with the physical requirements of charge and color neutrality, the
pairing pattern can be quite complex involving phases with gapless
modes for certain quark quasiparticles~\cite{KR1,KR2,Shovy2}. This
opens up the possibility of a direct urca process that is not
exponentially penalized even though the gap is non-zero, and the
neutrino emissivity from gapless CFL matter (gCFL) was calculated
in~\cite{KR04}. To facilitate a comparison of the rates, we choose (as
in~\cite{KR04}), $\mu_q=500$ MeV, $m_s^2/\mu_q=50$, and a gap
$\Delta=25$ MeV. From Fig.\ref{cflcomp}, it is clear that this is a
powerful channel for neutrino emission at all temperatures relevant to
the long-term cooling of the star. It is a sensitive function of
$m_s^2/\mu$, but the values used here are likely close to those in
nature. Its effect on overall cooling only shows up in the
photon-dominated cooling epoch however, due to the large specific heat
of the gapless phase~\cite{KR04}.

\vskip 0.2cm

Several other color superconducting phases at intermediate densities relevant to neutron stars have been considered in the literature~\cite{Schmitt1,Huang,Alfie} which we do not address individually here. The question as to which is the preferred stable state of quark matter at intermediate densities and physical strange quark mass remains an open one at this time. We may mention here that a crystalline (LOFF) phase may exist at large values of the strange quark mass~\cite{Casal,Bower} and the neutrino emission from such a phase remains to be computed within the framework of the low-energy effective Lagrangian for this phase~\cite{Nardulli}. 

\vskip 0.2cm

The enhancement of neutrino emission due to the inclusion of soft exchanges in electron-electron collisions can also be extended to quark-quark collisions, though the issue is much more subtle. Neutrino bremsstrahlung rates from quark-quark collisions have been previously computed in~\cite{Iwamoto,Price,Duncan83}, although without damping in the transverse channel taken into account. A treatment of quark-quark scattering and transport properties in a degenerate quark-gluon plasma including this effect was undertaken in~\cite{HP93}. Having performed an explicit computation for electron-electron scattering, we may expect a simple extension to neutrino bremsstrahlung in quark-quark scatterings, and for different quark phases. For ungapped quark matter with chemical potential $\mu$ for a particular flavor, the emissivity from eqn.(\ref{bremss1}) applies to scattering between like-flavor quarks, with an extra factor of 4/3 from color-counting and averaging, and with the replacements $e\rightarrow g$ for the coupling constant, $\mu_e\rightarrow\mu_q$, $c_{\rm eff}^2\rightarrow c_q^2$. The weak couplings for quarks $c_q$ are listed in Table 1 of~\cite{JP1}. The Debye mass is $m_D^2=\alpha_c\mu_q^2/\pi$. However, as pointed out in section II, at energies (temperatures) $\omega\sim m~{\rm e}^{-9\pi/4\alpha_s}\sim 10{\rm s~of~keVs}$, non-Fermi liquid effects can render a perturbative analysis invalid, leading to logarithmic enhancements in the quark self-energy and the neutrino rate in ungapped quark matter~\cite{SS1,SS2}. Already for $T_9<0.1$, non-Fermi liquid effects are expected to set in. Since we integrate over all exchange energies, a more accurate evaluation of bremsstrahlung neutrino rates should include the non-Fermi liquid effects. In this respect, the neutrino rate computed here will be somewhat modified for quark matter, but only very slightly so for a QED plasma. Further complications arise if quark matter is superconducting. For gapped quark matter, such enhancements as discussed here need only be considered when an ungapped quark species exists, and interacts via a gluon whose Meissner mass (but not the Debye mass) vanishes. In the CFL phase, all quarks are gapped, and all transverse gluons acquire a Meissner mass~\cite{Rischke00}.~\footnote{The two-flavor superconductor, i.e, the 2SC phase admits ungapped quarks and unscreened gluons, but is energetically disfavored as a homogeneous phase in most studies~\cite{Alfie,Alf}.} In the gapless  phase, the issue of Meissner masses is far from clear~\cite{Shovy3,RC,Alfo}. However, a residual $U(1)$ symmetry in this phase implies that for a condensate that is neutral under the associated conserved charge (denoted $\tilde{Q}$), the Meissner mass vanishes. Since some quark species carry $\tilde{Q}$ charge however, the gCFL phase is not a $\tilde{Q}$ insulator, and the $\tilde{Q}$ photon (like the other gauge bosons in this phase) is likely screened. Finally, in the LOFF phase~\footnote{An alternate phase that can arise at intermediate density is the Overhauser phase, which like the LOFF phase, is anisotropic; however it seems not to be favored energetically unless the number of colors is large~\cite{park} or the number of space dimensions is reduced~\cite{JZ}.}, whose appearance may be linked to the chromomagnetic instability of the gapless phases, the 5 gluons corresponding to the unbroken $SU(2)_c$ group acquire Debye masses due to the existence of blocking regions on the Fermi sphere, while the Meissner masses vanish~\cite{Fabio}. Since $u$ and $d$ quarks of one color (``blue'') as well as strange quarks of all colors (presumably too heavy to participate in pairing) remain unpaired, unscreened gluon exchange among them can lead to enhanced neutrino bremsstrahlung emissivities, although the dominant neutrino emission process would be the modified urca among ungapped light quarks (eg. $d+d\rightarrow d+u+e^-+\bar{\nu}_e$)~\footnote{The quark direct urca process $d\rightarrow u+e^-+\bar{\nu}_e$ would be suppressed due to the mismatch $\delta\mu=\mu_d-\mu_u$ being large in this phase.}.    

\vskip 0.2cm

\section{Comparison to neutrino rates in black-hole accretion tori}
         \label{sec:BH}

Extreme conditions in which neutrino emission is crucial for the
system evolution are also found in accretion tori around black-holes.
Such tori are essential ingredients of most Gamma-Ray Burst models,
either in the collapsar scenario for long bursts \cite{W93} or in the
neutron star-neutron star or neutron star-black hole merger scenarios
for short bursts (see, e.g., \cite{LRRP04} and \cite{LRRP05}).  With
central densities above $10^{12}$ g cm$^{-3}$ and temperatures in the
range of 1 to 10 MeV, nucleons form an almost ideal gas while
electrons are mildly degenerate.  Considering, for illustration,
comparable fractions of neutrons and protons, one has for the electron
chemical potential $\mu \sim 10 \; \rho_{10}^{1/3}$ MeV.  In the outer
part of the torus, electrons are non-degenerate and thus positrons are
present.

\vskip 0.2cm

A reliable comparison of the various neutrino emission processes can
only be performed through numerical calculations due to the wide range
of physical conditions present in such tori and their rapid time
variation, of the order of tens of milliseconds.  We only present here
some estimates based on the physical parameters that were employed in
the figures of~\cite{LRRP05}.  The dominant neutrino cooling process
in these models of accretion tori is electron capture on protons
\cite{KM02}, i.e., $e^- + p \rightarrow n + \nu_e$, which in the
degenerate electron limit, has a total emissivity
\beq
Q_{\mathrm{Capt}}^{\mathrm{Deg}} = 4.2 \times
10^{31} \; \mu_{10}^ 9 \; {\rm erg~cm}^{-3}{~\rm s}^{-1}\,, 
\eeq 
while eqn.(\ref{bremss2}) gives
\beq 
Q_{\rm ee-Br}=4.6\times 10^{30} \; T_{11}^7 \; \mu_{10}^2 \; \tilde{N}(\alpha) ~{\rm
erg~cm}^{-3}{~\rm s}^{-1}\,,
\eeq 
showing that neutrino bremsstrahlung from e-e scattering can make a significant contribution in the ``low'' density and temperature regions of the torus, and even become the dominant emission mechanism.
Maybe more important than energy loss is the production of antineutrinos since it is often considered that the jet of the GRB is triggered by pair production from $\nu$-$\overline{\nu}$ annihilation along the black-hole rotation axis.
Neutrinos are copiously produced by electron capture but antineutrinos come
essentially from the outer part of the torus, where $T \sim 10^{10}$ K and 
$\rho \leq 10^{10}$ g cm$^{-3}$, and are produced by
 bremmsstrahlung in nucleon-nucleon collisions, i.e., 
$N+N' \rightarrow N+N'+\nu+\overline{\nu}$, with a rate \cite{HR98} 
\beq
Q_{\mathrm{NN-Br}} = 4.7 \times 10^{21} \, T_{10}^{5.5} \, \rho_{10}^{2}\, 
{\rm
erg~cm}^{-3}{~\rm s}^{-1}\,,
\label{Eq:Q-NN-Br}
\eeq
where $\rho_{10}$ is the density in $10^{10}$ g/cc. Other important
contributions arise from the outermost regions, at densities $\rho \ll
10^{10}$ g cm$^{-3}$ where electrons are non-degenerate and positrons
are also present, from the process of pair annihilation, $e^- + e^+
\rightarrow \nu + \overline{\nu}$, with a rate
\beq
Q_{\mathrm{Pair}} = 4.8 \times 10^{24} \; T_{10}^9 \;
 {\rm
erg~cm}^{-3}{~\rm s}^{-1}\,,
\eeq
and positron capture on neutrons, $n + e^+ \rightarrow p + \overline{\nu}_e$,
with a rate
\beq
Q_{\mathrm{Cap}} = 4.6 \times 10^{24} \; T_{10}^6 \; \rho_{7} \;
{\rm
erg~cm}^{-3}{~\rm s}^{-1}\,,
\eeq
However, neutrinos from e-e scattering which, in contradistinction to
processes involving positrons, come from within the whole torus volume, can
be expected to make by far the dominant contribution to antineutrino
production. Inclusion of this process in future numerical simulations
will allow an assessment of its importance.

\section{Comparison to neutrino rates in neutron stars}
         \label{sec:NS}

Neutrino pair bremsstrahlung in electron-electron collisions can
also be important in neutron stars and we discuss this in the
present section.  In the star's crust, the major neutrino emission
processes are pair bremsstrahlung processes for electron-ion
scattering while in the neutron $^1$S$_0$ superfluid regime, it is
neutrino pair emission by breaking and formation of Cooper pairs at a
temperature just below $T_c$, the critical temperature for pairing.  The
Cooper pair process is by far the dominant one but only acts
efficiently in a narrow density region where the temperature $T \sim 0.2 - 0.9 \; T_c$, because of the density dependence of $T_c$, 
while the e-ion and e-e bremsstrahlung processes act in the whole
crust.  We plot in the left panel of Fig.~\ref{Fig:NS} these last two
processes: the emissivity of neutrinos in e-e collisions can dominate
only at temperatures $\leq 10^8$ K and the highest density regime
where the ``pasta'' phase is present.  One must emphasize that this
``pasta'' region actually comprises almost half of the mass of the
crust and, moreover, neutrino bremsstrahlung in e-ion collisions is
expected to be actually much weaker than shown in the figure
\cite{KPPTY99}, by up to two orders of magnitude, due to the lower
dimensionality of the nuclei (which are rods and plates instead of
spheres).

In the core of the neutron star, the modified Urca and nucleon
bremsstrahlung processes are usually considered as the main neutrino
emission processes in the absence of any enhanced neutrino emission
process such as, e.g., the direct Urca. We plot in the right panel of
Fig.~\ref{Fig:NS} the emissivities of the modified Urca process as
well as the bremsstrahlung rates in p-p (with rates taken from
\cite{YL95}) and e-e collisions: naturally, the modified Urca process
is the dominant one.  However, in regions where neutrons are
superfluid and $T\ll T_c$ the modified Urca process is strongly
suppressed, as is neutrino pair emission by the Cooper pair formation
process and n-n and n-p associated bremsstrahlung: in such a case only
the bremsstrahlung from p-p and e-e scatterings is effective and the
latter one is dominant.  Notice that in case of proton pairing, the
bremsstrahlung rate from p-p scattering will become suppressed but
that from e-e will also be significantly reduced because in a superconductor
 static magnetic fields are screened by the Meissner effect. Photons
acquire a Meissner mass and electromagnetic interactions are
thus fully screened instead of having their transverse part only
Landau-damped. In such a case the calculation of \cite{Haensel}
should be applicable for the bremsstrahlung emissivity from e-e
scattering, since these authors assumed screening of the whole
propagator by a Debye mass $m_D$ (instead of $m_D$ and $m_M$). Again,
this process turns out to be the dominant neutrino emission process,
although quite inefficient.  In the outer part of the core, protons
are certainly paired with a $T_c$ of the order of $3 \times 10^9$ K at
most \cite{AWP91}, while they are likely to be in a normal state at
higher densities; the possibility that neutrons may pair at these
densities is still very controversial and predictions range from
possibly quite high values of $T_c$ \cite{BEEHS98} to almost zero
\cite{SF04}.

\begin{figure}[t]
\bce
\epsfig{file=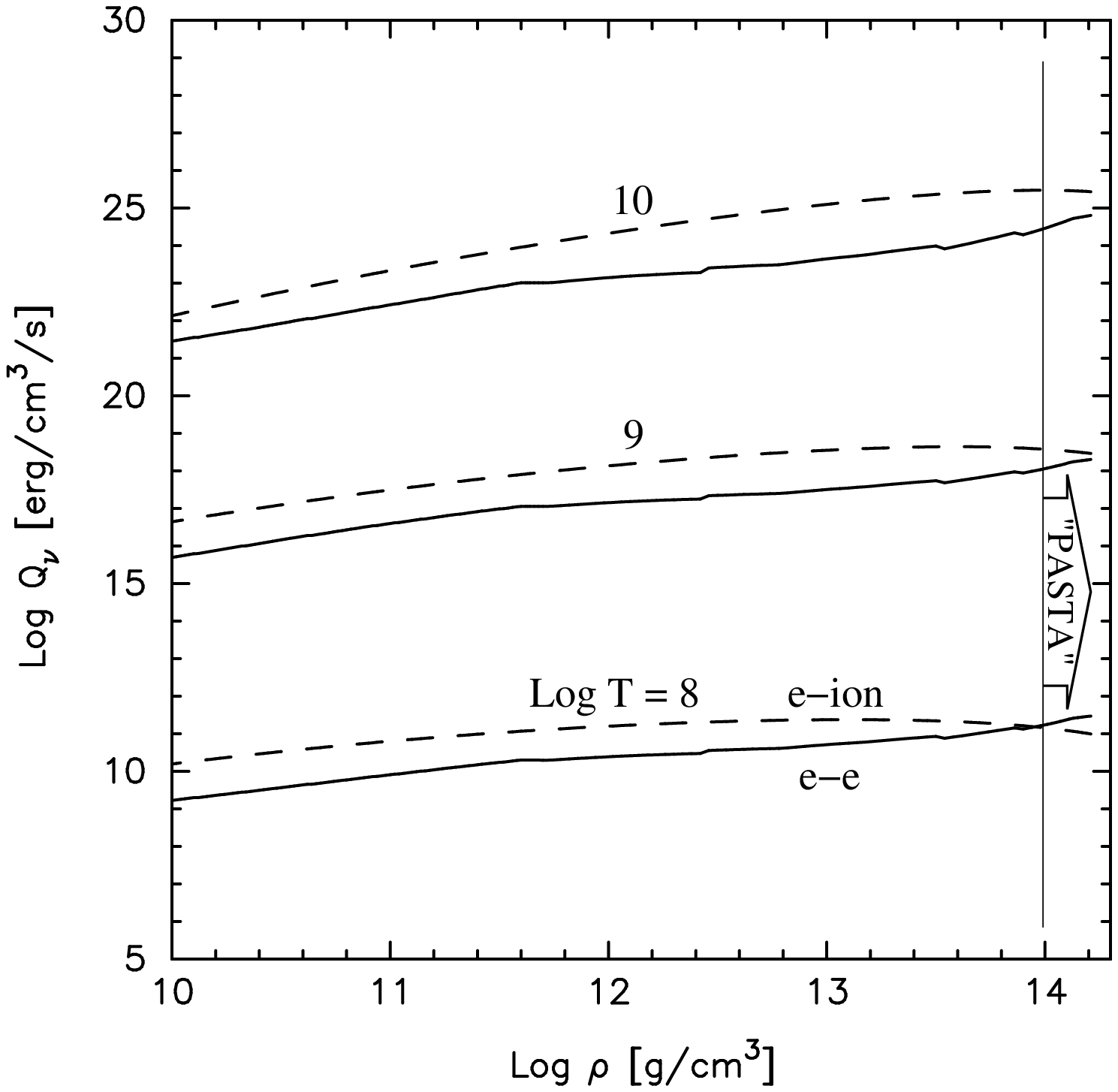,scale=0.5}
\epsfig{file=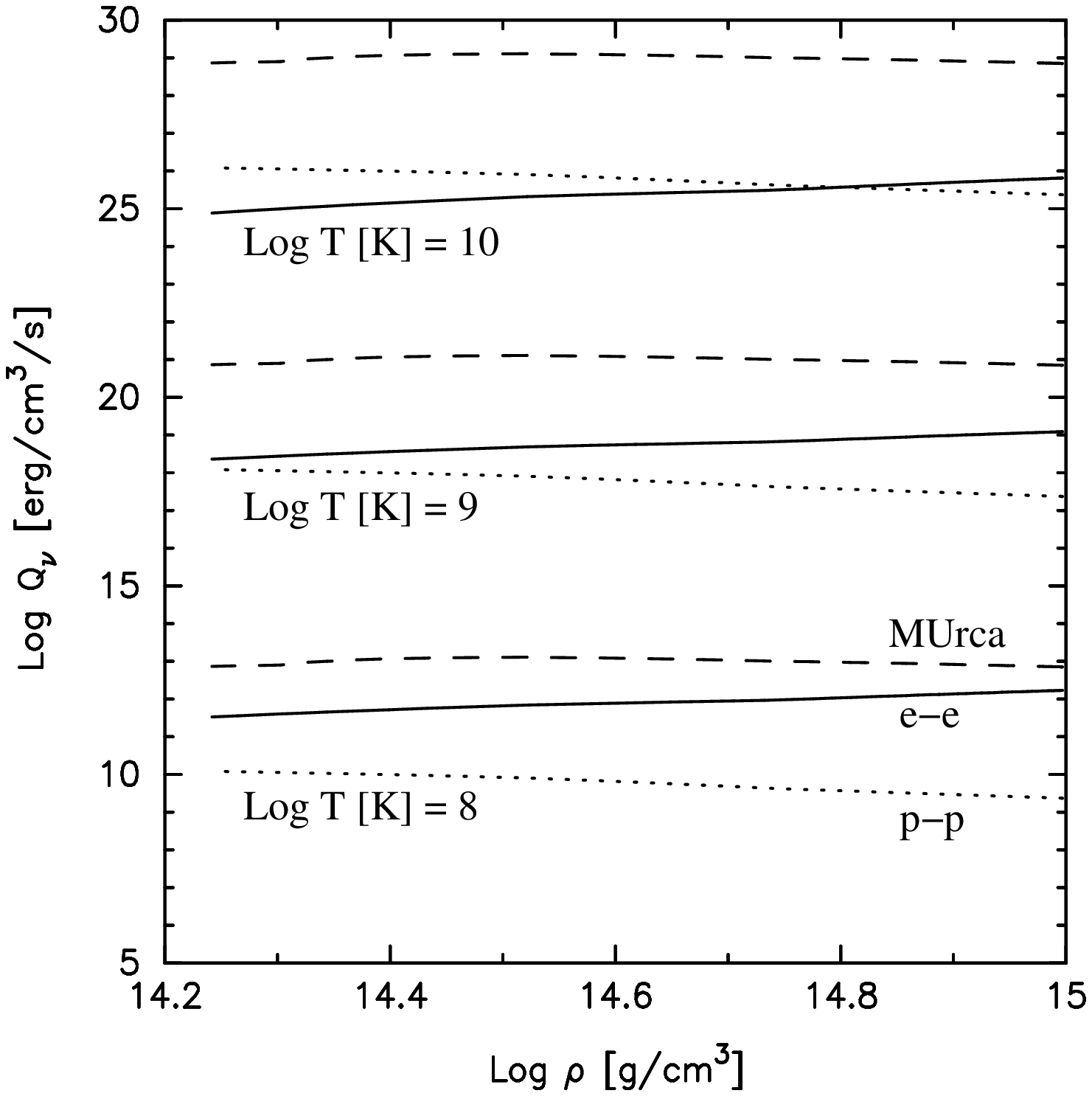,scale=0.5}
\ece
\caption{Relevance of the e-e bremsstrahlung process in neutron stars.
         Left panel: comparison with the electron-ion bremsstrahlung neutrino emission
                     in the crust, at three typical temperatures.
         Right panel: comparison with the modified Urca (``MUrca''),
                      sum of the neutron and proton branches, and the 
                      proton-proton bremsstrahlung process 
                      in the core at the same temperature.}                      
\label{Fig:NS}
\end{figure}

\section{Conclusions and Outlook}
         \label{sec:Conclusion}

We have performed an explicit computation of the neutrino emissivity
from neutrino bremsstrahlung in electron-electron collisions in a
relativistic degenerate plasma within an astrophysical context. This
process has been studied before in the context of neutrino
cooling from neutron star crusts, and in gapped superfluids. In this
work, we have re-investigated this process in different astrophysical
settings, with closer attention to the dynamics of the screened photon
propagator that mediates electron-electron collisions. Compared to
previous results~\cite{Haensel}, we find an enhancement when 
Landau damping in the transverse part of the photon propagator is
taken into account. If we assume simple screening in both the
longitudinal and transverse channels of photon exchange, we recover
the results of~\cite{Haensel}. Our principal results are
eqn.(\ref{bremss2}) and the comparisons in Fig.\ref{figure2},
Fig.\ref{cflcomp} and Fig.\ref{Fig:NS}. The inclusion of Landau damping
implies a $\mu_e^2~T^7$ dependence, instead of the $\mu_e~T^8$
dependence as expected from screening arguments. This leads to an
enhancement in the neutrino rate which increases with decreasing
temperature, suggesting that e-e bremstrahlung is indeed an efficient
cooling process in a certain window of temperature and density. This
conclusion could be of particular importance to cooling of strange
stars when the interior of the strange star is in a color
superconducting phase, such as the CFL phase, which has only gapped
quark excitations. The neutrino emission from the interior is then
suppressed by several orders of magnitude as the temperature falls
below the critical temperature for pairing ($T_c\sim 50$ MeV). The
overall cooling of the star can be affected due to surface effects as
shown in~\cite{Page02,JPPG}. Inclusion of the enhanced neutrino rate in the
computations of~\cite{Page02} would be a natural extension of this
work, and taken in conjunction with the photon cooling from the
strange star surface (which can also influence the light curve at
early times) serve to present a more complete picture of strange star
cooling.

\vskip 0.2cm

The quark matter underlying the strange star surface can exist in any
one of many phases, and neutrino emission rates have been computed in
only a few of them. Of particular interest are the gapless and
crystalline phases, which will lead to temperature dependences that
are modified from the usual forms in ungapped quark matter. These
phases also have unique dispersion relations for certain quark
quasiparticles, and consequently, a specific heat per unit volume that
is also different from ungapped quark matter. These two factors imply
a change in the stellar cooling curve that can be confronted by
observations. While we have focused on computing a particular process, our ultimate goal in this direction is to provide more accurate
neutrino rates, including the possibility of various quark phases, in
order to make accurate predictions for the cooling history of a
strange star. This task is complicated by the appearance of non-Fermi
liquid behaviour, and the breakdown of a perturbative treatment at
temperatures relevant to the long-term cooling of quark matter in the
stellar interior. 

\vskip 0.2cm

Finally, we have established the relevance of neutrino bremsstrahlung in e-e collisions for black-hole accretion tori and crustal layers of ordinary neutron stars. The GRB jet may be triggered by $\nu\bar{\nu}$ annihilation, while its subsequent cooling also depends on the neutrino flux, and the e-e process is important for both. Neutron star cooling can also be affected by this process at low temperatures in the pasta-phase of the crust or when neutrons are superfluid in the core. Our results thus find wide-ranging applications for neutron stars, GRB's and strange stars. Increasing data on neutron star cooling, GRB 
afterglow signatures and theoretical advances on observable signatures of strange stars 
\cite{PC05} all suggest that a renewed effort to pin down neutrino rates for astrophysics with improved accuracy is both timely and interesting.


\section*{Acknowledgments}
P.J. is supported by the Department of Energy, Office of Nuclear
Physics, contract no. W-31-109-ENG-38. He thanks Guy Moore and Thomas
Sch\"afer for clarifying discussions. The work of D.P. is partially
supported by grants from UNAM-DGAPA (PAPIIT-IN112502) and Conacyt
(36632-E). C.G. is supported in part by the Natural Sciences and
Engineering Research Council of Canada and in part by the Fonds Nature
and Technologies of Quebec.\rm


\appendix 
\section{$k$-Integration}
In eqn.(\ref{kintegral}), a change of variable to $x=\omega/k$ gives
\beqy
I_k&=&\frac{1}{\omega^3}\int_{a}^{\sim\omega/l_1}dx\frac{x^2}{\left[(1-x^2)^2+b^2x^6\right]}\left[1-\left\{\frac{a(1-x^2)+x^2}{x}\right\}^2\right]\quad ; \label{kintegral2}\\
a&=&\frac{\omega}{2\mu_e};\quad b=\frac{\pi}{4}\left(\frac{m_D}{\omega}\right)^2\quad.
\eeqy
For the physical conditions of interest, $a\ll 1$ and $b\gg 1$, which enables us to use a much simpler form
\beq
I_k=\frac{1}{\omega^3}\int_{a}^{\sim\omega/l_1}dx\frac{x^2}{\left[1+b^2x^6\right]}\quad .\label{kintegral3}
\eeq 
The integrand is not very different than in eqn.(\ref{kintegral2}) for the entire range of integration (note that $l_1\sim T$ so that $\omega/l_1\sim 1$), as evident from Fig.~\ref{figure4} below. 

\begin{figure}[!ht]
\bce
\epsfig{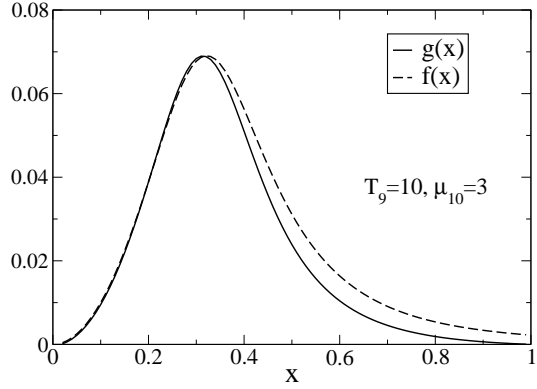}
\ece
\caption{The integrands from eqn.(\ref{kintegral2}) and eqn.(\ref{kintegral3}) labeled $g(x)$ and $f(x)$ respectively. The temperature is $T_9=10$ ($T=10^{10}$K) and $\mu_{10}=3$ ($\mu_e=30$ MeV).}
\label{figure4}
\end{figure}
The integral in eqn.(\ref{kintegral3}) is straightforward and yields
\beq
I_k\approx \frac{4}{3\pi m_D^2\omega}\left[\frac{\pi}{2}-{\rm tan}^{-1}\left(\frac{l_1^3}{\frac{\pi}{4}m_D^2\omega}\right)\right]\quad. 
\eeq

\section{Numerical evaluation of $N(\alpha)$}
The integral $I(\bar{\Omega})$ in eqn.(\ref{nalpha}) is
\beqy
I(\bar{\Omega})&=&\int_0^{\infty}d\bar{\omega}\frac{(\bar{\omega}+\bar{\Omega})}{{\rm sinh}\bar{\omega}{\rm sinh}(\bar{\omega}+\bar{\Omega})}\left[\frac{\pi}{2}-{\rm tan}^{-1}\left(\frac{4}{\pi}\frac{(\bar{\omega}+\bar{\Omega})^3}{\bar{\omega}\alpha^2}\right)\right]+\int_0^{\bar{\Omega}}d\bar{\omega}\frac{\omega}{{\rm sinh}\bar{\omega}{\rm sinh}(\bar{\omega}+\bar{\Omega})}\left[\frac{\pi}{2}-{\rm tan}^{-1}\left(\frac{4}{\pi}\frac{(\bar{\omega}+\bar{\Omega})^2}{\alpha^2}\right)\right]\nonumber\\
&+&\int_0^{\bar{\Omega}}d\bar{\omega}\frac{(\bar{\Omega}-\bar{\omega})}{{\rm sinh}\bar{\omega}{\rm sinh}(\bar{\Omega}-\bar{\omega})}\left[\frac{\pi}{2}-{\rm tan}^{-1}\left(\frac{4}{\pi}\frac{(\bar{\Omega}-\bar{\omega})^3}{\bar{\omega}\alpha^2}\right)\right]+\int_{2\bar{\Omega}}^{\infty}d\bar{\omega}\frac{(\bar{\Omega}-\bar{\omega})}{{\rm sinh}\bar{\omega}{\rm sinh}(\bar{\Omega}-\bar{\omega})}\left[\frac{\pi}{2}-{\rm tan}^{-1}\left(\frac{4}{\pi}\frac{(\bar{\omega}-\bar{\Omega})^3}{\bar{\omega}\alpha^2}\right)\right] \label{Iomega}
\eeqy
with
\beq
\bar{\omega}=\frac{\omega}{2T}\quad .
\eeq
This integral is well-behaved for all finite values of $\bar{\Omega}$,
even as $\bar{\omega}\rightarrow 0$. At $\bar{\Omega}=0$, the
integrand diverges at $\bar{\omega}=0$, but the overall emissivity is
clearly zero. Thus, a numerical evaluation of $N(\alpha)$ from
eqn.(\ref{nalpha}) with the above expression for $I(\Omega)$ is
feasible. The rapid exponential fall-off of the integrand with respect
to $\bar{\omega}$ and $\bar{\Omega}$ implies a rapid convergence and
high degree of accuracy in the final result. The integral $N(\alpha)$
is evaluated by Gauss-Legendre quadratures and the normalized
$\tilde{N}$ that appears in eqn.~(\ref{bremss2}), is plotted in
Fig.\ref{figure5}.

\newpage

\begin{flushleft}

\end{flushleft}
\end{document}